# A Proposed Framework for Development of a Visualizer Based on Memory Transfer Language (MTL)

Ally S. Nyamawe

*The University of Dodoma, College of Informatics and Virtual Education*

*P.O.Box 490, Dodoma, Tanzania*

*Abstract*— Computer programming is among the fundamental aspects of computer science curriculum. Many students first introduced to introductory computer programming courses experience difficulties in learning and comprehending. Vast amount of researches have revealed that, generally programming courses are regarded as difficult and challenging and thus often have the highest dropout rates. Moreover, numerous researches have devoted in delivering new approaches and tools in enhancing the process of teaching and learning computer programming to novice programmers. One among the tools that have emerged to offer positive results is Program Visualization tool (Visualizer). Visualizers have shown remarkable contributions in facilitating novices to learn and comprehend computer programming. In addition to that, an approach to visualize codes execution, Memory Transfer Language (MTL), allows a novice to animate the code through paper and pencil mechanism without actively involving the machine. MTL depends on the concepts of RAM (Random Access Memory) to interpret the code line by line. Programming requires effort and special approach in the way it is learned and taught, thus this paper aimed at presenting a proposed framework for developing a visualizer that employs the use of MTL to enhance teaching and learning programming.

## I. INTRODUCTION

Computer programming lies in the core of computer science curriculum. Programming plays a very fundamental role in delivering software solutions. To program, one needs to have an ability to understand and analyse problem, and writing an algorithm that will solve the identified problem. Difficulties in teaching and learning programming have been enduring for years and now considered a universal problem. In the first place, the difficulties in teaching and learning programming were fuelled by the poor pedagogical approaches. In recent years, a substantial number of researchers have devoted in delivering newly pedagogical approaches and tools that will make learning programming an interesting endeavor. With the advances in multimedia technology, educational tools have been employing the use of audio, video and animation to create a more user friendly and interactive teaching and learning environment. For long, teaching a programming language in the classroom has been a challenge [1]. To enhance teaching and learning computer programming, assistive approaches and tools need to be evolved.

To visually and dynamically explain the concepts of computer programming, program visualization tools (Visualizers) were suggested. Visualizers use animated graphical representation to display line by line execution of the computer program. Rosminah et al [2] posit that, the main goal of a visualizer is to assist students in comprehending the dynamicity of the program through displaying aspects like values of variables, evaluation of statements, and changes in the program state in general. Furthermore [2] contended that, the hidden processes during the program run-time could also be explained visually by a visualizer. Generally, visualizers have so far offered positive results. In the research conducted by Rajala et al [3], through their visualizer named ViLLE, the authors concluded that, program visualization tool enhances student's learning regardless of previous programming experience. The positive results were also reported by Kasurinen et al [4]. Moreover, the research conducted by Mselle et al [5], on the effectiveness of MTL in learning programming, the results showed that MTL improves comprehension of the subject. In addition to that, Mselle [6] research findings suggested that, the use of RAM diagrams as a visualizer improves programming comprehension and programming skills.

### A. Overview of MTL

As defined by Mselle et al [5], Memory Transfer Language (MTL) is a language or device used by programmers to describe the impact of code-line on computer memory (RAM). The use of RAM blocks to demonstrate the effect of lines of code to a machine allows novices to grasp what actually is going on to a machine as results of program execution. MTL demonstrates the status of RAM blocks in three situations; before code execution, during code execution and after code execution. Throughout the process of executing a program, RAM blocks are dynamically changed to reflect the effect of a line in execution. A step by step code execution allows a novice to reason and predict the final results which in turn enhances comprehension. MTL takes novice into board during execution process, and therefore the power of novice to control the machine is not deprived. Mselle et al [7] posit that, it is proved that MTL is a language to learn programming which allows novices to develop their coding skills through practicing and two-way-thinking approach.

*B. Program Visualization Tools*

Numerous program visualization tools have been developed and deployed. Though the rationale behind all visualizers is almost the same, each one has come up with unique features and functionalities and distinct ways of helping novices to program. Mutua et al [8] posit that, visualization tools have been developed to supplement the learning process; in that regard various forms of teaching aids, models and software systems are deployed in enhancing the learning process. Different visualizers for different programming languages such as C, C++ and JAVA are available today. Kasurinen et al [4], deployed a program visualization tool (Turtlet) to enhance student motivation and interest towards programming in the introductory programming course by applying visualization tool to lecture demonstrations. Alice, a 3-D interactive graphics programming environment developed with a goal to make it easy for novices to develop interesting 3-D environments and to explore the new medium of interactive 3-D graphics. Alice serves as a good programming language for the novice programmers as they can follow up and see how their animated programs run [9]. Some other visualizers developed include BlueJ, Jeliot, JPie and Scratch.

## II. PROPOSED FRAMEWORK

*A. Overview*

In this section we present a proposed framework that could be used in developing a visualizer based on the MTL framework. A visualizer should be designed to aid a novice programmer to visualize a program as it is being executed by the machine. A visualizer is expected to allow a novice to construct his/her program of choice by either using the built in controls to automatically insert codes or coding from the scratch. Once the code is entered, the execution of the lines of code can be triggered and execution begins. A visualizer highlights (in colour) each and every line it executes at a particular point in time and displays the effect to a machine over the RAM blocks. A visualizer is anticipated to allow a novice to select the mode of program execution. The first mode is line by line execution, where the tool reads one line at a time, highlights a line in colour, and finally shows the code effect over the RAM blocks. In this mode a novice's intervention is needed to allow a visualizer to execute the next line. The second mode of execution allows for a complete run, where the tool executes the program from the beginning to the end while showing the code effect over the RAM blocks. A visualizer shall provide user friendly error messages that provide suggestions and guide a novice on how to rectify the reported errors. In this paper we will demonstrate examples of codes/programs in VB .Net, however in implementation; one may choose whatever language of choice.

*B. A Visualizer's Layout*

As depicted in Figure 1, a visualizer is proposed to have three main parts:

1. The controls window: constitutes controls for adding statements automatically in the code window (A place where a program is displayed). The controls are for "Declaration", "Assignment", "Data Input", "Data Output", "Condition Statement", "Looping Statement" and "Insert text/statement".

   This approach allows novices to create simple programs without memorizing the syntax of a programming language in use, which in turn reduces cognitive load for beginners. In addition to that, novices are not vulnerable in making syntactical errors as everything is taken care by a visualizer. A novice is simply required to click the control of the functionality to be performed and follow instructions. However, a visualizer should leave the room for the conversant novices to write their program in the code window right from the scratch without using the built in controls to automatically generate codes.

2. A code window: where the program statements are displayed. After the novice selects a particular control to insert a statement, such as for the variable declaration or data feeding. The corresponding code is automatically generated and displayed in the code window. A code window also allows a novice to manually type in a program. When a program is in execution, a line currently executed is highlighted in colour to visually demonstrate a step by step execution of a program and affect the corresponding RAM block accordingly.

3. RAM diagrams: have three memory blocks. The first part is the block of RAM before program execution. This block is used to display the status of RAM before execution of any line of code. The block is displayed and seen to contain nothing. The second RAM block, is the one which shows the status of RAM after variable declaration. Once a line for declaring a variable is executed it affects the block of RAM by showing that, the memory space is reserved for each declared variable. The third RAM block displays the status of RAM after values are assigned to their corresponding variables. The third block displays the name of a variable and the memory location containing the value that has been assigned to a variable.

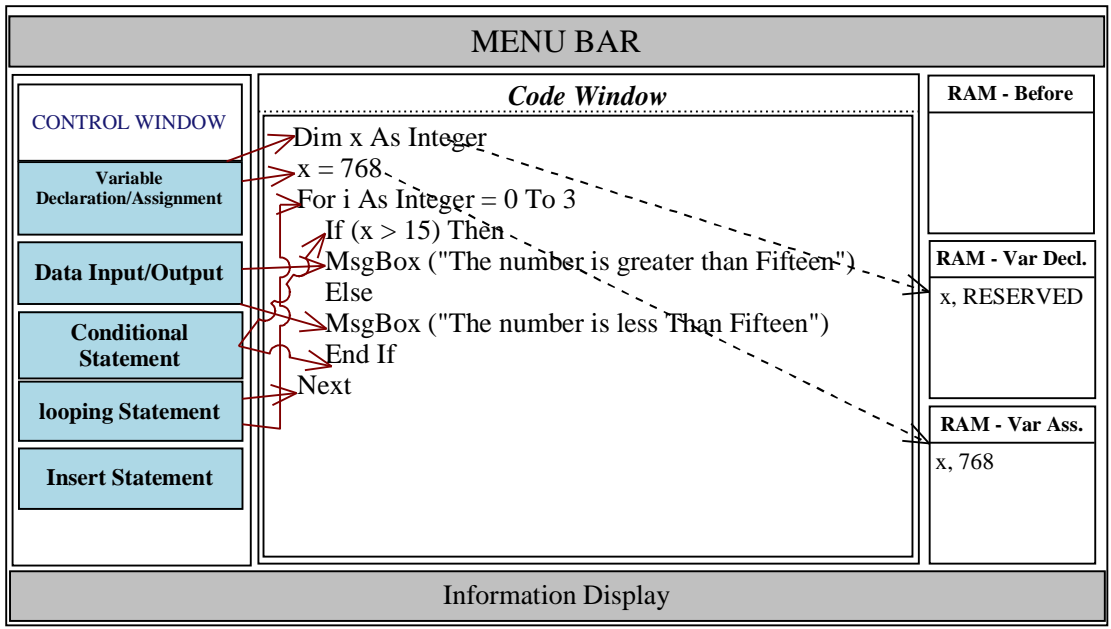

Fig. 1 The Proposed Layout of a Visualizer Interface

*C. Program Execution*

A visualizer is required to clearly demonstrate through animations the execution of the program and their corresponding effects to the machine memory. In this part we present an execution of a sample program written in VB .Net to just provide a general overview on how a visualizer should behave during execution. A program under execution as shown in Table 1, asks a user to input two numbers which are stored in array and output their sum.

Table 1: A Sample Program under Execution

```
1. Dim sum As Integer
2. sum = 0
3. Dim num (1) As Integer
4. For i As Integer = 0 To 1
5.   num(i) = InputBox("Input number" + i)
6.   sum= sum + num(i)
7. Next
8. MsgBox("The sum of numbers is" + sum)
```

**Execution of line 1:**

```
1. Dim sum As Integer
2. sum = 0
3. Dim num (1) As Integer
……..
```

Declaration of a variable named *sum*.

Fig. 2 A code window displaying line 1 in execution

Before program execution, the RAM diagram is displayed empty. As depicted in Figure 2, once the execution is started, a line in execution is highlighted or displayed in a different way comparing to the rest so as to draw the attention of a novice and provide details on what a line is all about. In line 1, a variable named *sum* is declared. A RAM block will be displayed to show that a memory location that will store a value to be assigned to *sum* is reserved. As depicted in Figure 3, RAM block will be accompanied by details to tell a novice what has happened in a machine as result of a line execution.

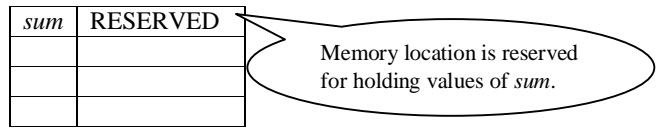

Fig. 3 The status of a RAM block after execution of line 1

**Execution of line 2:**

```
1. Dim sum As Integer
2. sum = 0
3. Dim num (1) As Integer
……..
```

Assigning 0 to a variable *sum*

Fig. 4 A code window displaying line 2 in execution

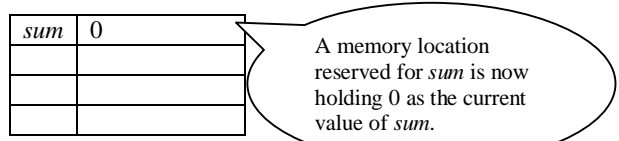

Fig. 5 The status of a RAM block after execution of line 2

**Execution of line 3:**

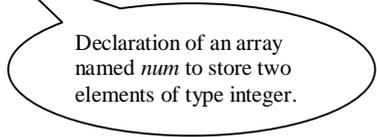

Fig. 6 A code window displaying line 3 in execution

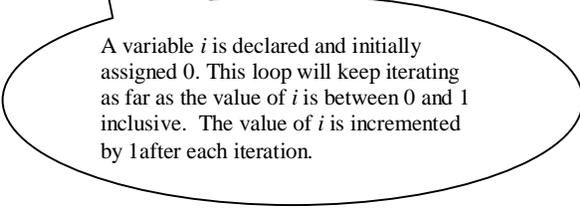

Fig. 7 The status of a RAM block after execution of line 3

**Execution of line 4:**

```
........
3. Dim num (1) As Integer
4. For i As Integer = 0 To 1
........
```

A variable *i* is declared and initially assigned 0. This loop will keep iterating as far as the value of *i* is between 0 and 1 inclusive. The value of *i* is incremented by 1 after each iteration.

Fig. 8 A code window displaying line 4 in execution

The effect of execution of line 4 to machine memory is as depicted in Figure 9.

| sum | 0 |
| num(0) | RESERVED |
| num(1) | RESERVED |
| i | 0 |

A memory location holding 0 as the current value of *i*

Fig. 9 The status of a RAM block after execution of line 4

As far as the loop condition (if the value of *i* is between 0 and 1 inclusive) is true, the statements in the loop body (line 5 and 6) will be executed iteratively until the loop condition becomes false. A visualizer should therefore demonstrate the cycle of execution during looping.

**Execution of line 5:**

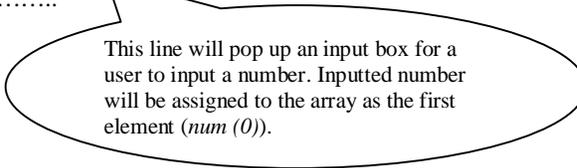

Fig. 10 A code window displaying line 5 in execution

Suppose a user inputs number *409*, the resulting RAM diagram is as depicted in Figure 11.

| sum | 0 |
| num(0) | 409 |
| num(1) | RESERVED |
| i | 0 |

A memory location reserved for the first element of array *num* is now holding number 409.

Fig. 11 The status of a RAM block after execution of line 5

**Execution of line 6:**

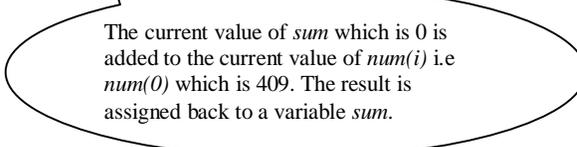

Fig. 12 A code window displaying line 6 in execution

| sum | 409 |
| num(0) | 409 |
| num(1) | RESERVED |
| i | 0 |

A memory location holding 409 as the current value of *sum*.

Fig. 13 The status of a RAM block after execution of line 6

After all statements in the loop body are executed, the *next* line (line 7) will be executed. This command returns control or directs the execution back to the beginning of the loop (line 4) until the loop condition becomes false. Therefore, line 5 and 6 will be executed for the second iteration.

**Execution of line 7:**

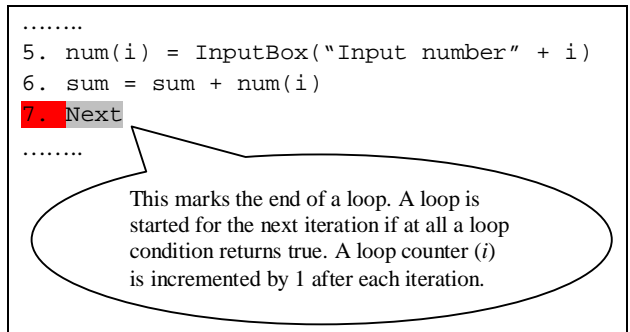

Fig. 14 A code window displaying line 7 in execution

Execution of line 7 causes a loop counter to be incremented by 1 unless otherwise it is explicitly defined. The effect of line 7 to a RAM is as depicted in Figure 15.

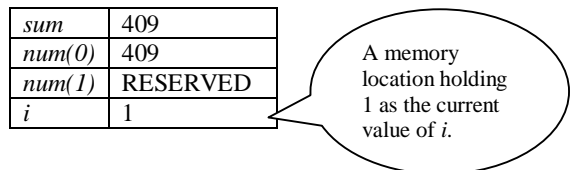

Fig. 15 The status of a RAM block after execution of line 7

All loop iterations shall be clearly demonstrated by a visualizer. Therefore execution of line 4 to 7 will be repeated as demonstrated from Figure 8 to 15. But this time there would be some few changes depending on the user inputs.

A visualizer shall explicitly exhibit each line in execution and corresponding effects to a machine memory. Whatever way could be used for illustration purpose, such as graphical representation, animation or audio. On top of that, a novice shall have ability to control a visualizer such as replaying execution, pauses execution and selects speed of execution.

## III. CONCLUSION

With the tool that provides program visualization, the novices are able to visualize what is going on in the machine. MTL provides the visualization of code during execution, showing each and every line with its effect on the RAM. This enables a novice to understand the effect of every line of code as it is executed by the machine. MTL is the current approach which transfers all authority to the novice programmer. Moreover, MTL has been tested in live classes and yielded positive results. Further researches are required to come up with tools that employ a framework of MTL to exploit its power in teaching and learning programming courses. In this paper we have just demonstrated some few examples on what a visualizer based on MTL should offer, but to develop a more comprehensive MTL based visualizer we encourage developers to consult articles in MTL for more details.